\documentclass[10pt]{article}
\usepackage{graphicx}

\def\Title#1{\begin{center} {\Large #1 } \end{center}}
\def\Author#1{\begin{center}{ \sc #1} \end{center}}
\def\Address#1{\begin{center}{ \it #1} \end{center}}

\newcommand\pubblock{\rightline{\begin{tabular}{l} Proceedings of the Fifth Annual LHCP\\ \pubnumber\\
         \pubdate  \end{tabular}}}

\newenvironment{Abstract}{\begin{quotation} \begin{center} 
             \large ABSTRACT \end{center}\bigskip 
      \begin{center}\begin{large}}{\end{large}\end{center} \end{quotation}}

\newenvironment{Presented}{\begin{quotation} \begin{center} 
             PRESENTED AT\end{center}\bigskip 
      \begin{center}\begin{large}}{\end{large}\end{center} \end{quotation}}

\def\beq{\begin{equation}}
\def\eeq#1{\label{#1}\end{equation}}
\def\eeqn{\end{equation}}

\def\beqa{\begin{eqnarray}}
\def\eeqa#1{\label{#1}\end{eqnarray}}
\def\eeqan{\end{eqnarray}}

\let\bar=\overbar

\def\Dslash{\not{\hbox{\kern-4pt $D$}}}
\def\dslash{\not{\hbox{\kern-2pt $\del$}}}

\def\msb{{\bar{\ssstyle M \kern -1pt S}}}

\usepackage{color}

\newcommand\pubnumber{ CMS-CR-2017-202 }

\newcommand\pubdate{\today}

\def\affiliation{
On behalf of the CMS Collaboration, \\
University of Wisconsin-Madison, Madison, United States of America}

\begin{document}

\large
\begin{titlepage}
\pubblock

\vfill
\Title{  Observation of Higgs boson decays to $\tau$ lepton pairs  }
\vfill

\Author{ Cecile Caillol }
\Address{\affiliation}
\vfill
\begin{Abstract}

 A search for Higgs boson decays to $\tau$ leptons is performed using events recorded in proton-proton collisions by the
CMS experiment at the LHC at a center-of-mass energy of 13 TeV. The data
set corresponds to an integrated luminosity of 35.9 fb$^{-1}$. An excess
of events is observed over the expected background prediction with a significance of
4.9 standard deviations, to be compared to an expected significance of 4.7 standard
deviations.  

\end{Abstract}
\vfill

\begin{Presented}
The Fifth Annual Conference\\
 on Large Hadron Collider Physics \\
Shanghai Jiao Tong University, Shanghai, China\\ 
May 15-20, 2017
\end{Presented}
\vfill
\end{titlepage}
\def\thefootnote{\fnsymbol{footnote}}
\setcounter{footnote}{0}
%

\normalsize 


\section{Introduction}

The discovery by the CMS and ATLAS Collaborations of a new particle with mass of 125 GeV compatible with the Higgs boson of the standard model (SM)~\cite{Aad:2012tfa,Chatrchyan:2012ufa}, in 2012, was led by the sensitivity of the bosonic decay channels. To validate the SM mechanism for the generation of fermion masses, it is crucial to demonstrate that the Higgs boson couples to fermions, and that the strength of these couplings is proportional to the fermion mass. \\

Among the leptonic decay channels, the branching fraction of $H\to\tau\tau$ is by far larger than those of $H\to\mu\mu$ or $H\to ee$. Although the branching fraction of $H\rightarrow bb$ is about ten times larger, the $H\to\tau\tau$ channel is more sensitive because of a better signal-to-background ratio.\\

The analysis~\cite{HIG-16-043} is based on 35.9 fb$^{-1}$ of proton-proton data collected by the CMS experiment at a center-of-mass energy of 13 TeV. All $\tau\tau$ final states are studied, except those with two muons or two electrons because of low branching fractions. In the following, $\tau$ leptons reconstructed in their hadronic decay modes are denoted as $\tau_h$.

\section{Event selection and categorization}

Events are selected using different triggers depending on the $\tau\tau$ decay mode: triggers require one electron and one muon objects in the $e\mu$ final state, one electron object in the $e\tau_h$ final state, one muon or one muon and one $\tau_h$ objects in the $\mu\tau_h$ final state, and two $\tau_h$ objects in the $\tau_h\tau_h$ final state. In the $e\tau_h$ and $\mu\tau_h$ final states, the $\tau_h$ candidate is required to have $p_T$ greater than 30 GeV; this threshold is raised to 50 and 40 GeV in the $\tau_h\tau_h$ channel for the leading and subleading $\tau_h$, respectively. The muon in the $\mu\tau_h$ final state is required to have $p_T>20$ GeV, and the electron in the $e\tau_h$ channel $p_T>26$ GeV. In the $e\mu$ final state, the leading lepton should have $p_T>24$ GeV, and the $p_T$ threshold for the subleading lepton is 15 GeV if it is a muon, or 23 GeV if it is an electron.\\

The W+jets background in the $e\tau_h$ and $\mu\tau_h$ final states is reduced by selecting only events with a transverse mass between the light lepton and the transverse missing momentum less than 50 GeV. In the $e\mu$ channel, the $t\bar{t}$ background is reduced by requiring $p_{\zeta} - 0.85 p_{vis} > -35$ or $-10$ GeV depending on the category, where $p_\zeta$ is the component of the missing transverse momentum  along the bisector of the transverse momenta of the two leptons and $p_{vis}$ is the sum of the components of the lepton transverse momenta along the same direction.\\

The events are separated into three categories:
\begin{enumerate}
\item 0-jet. This category contains events that do not have any jet with $p_T>30$ GeV and $|\eta|<4.7$. It targets the gluon fusion production of Higgs bosons.
\item VBF. This category contains events with at least (or exactly depending on the final state) two jets with $p_T>30$ GeV and $|\eta|<4.7$. Conditions on the invariant mass of the jets and on their separation in the $\eta$ direction, on the $p_T$ of the leptons, or on the $p_T$ of the $\tau\tau$ system, are imposed and vary among final states. The VBF category targets events produced via vector boson fusion.
\item Boosted. This category contains all events that fail to pass the 0-jet and VBF selection criteria. It is called "boosted" because it targets Higgs bosons produced via gluon fusion, and recoiling against jets from initial state radiation.
\end{enumerate}

In every category the results are extracted from a fit to two-dimensional (2D) distributions built from two sensitive variables. One of the observables is always the mass of the $\tau\tau$ system, reconstructed from the visible decay products or using a more advanced likelihood method to recover the energy lost by the neutrinos in $\tau$ decays. In the 0-jet category, the second dimension is the $p_T$ of the muon for the $e\mu$ final state, or the $\tau_h$ reconstructed decay mode in the $e\tau_h$ and $\mu\tau_h$ final states. No second dimension is considered in the 0-jet category of the $\tau_h\tau_h$ final state. In the VBF category, the second dimension is the invariant mass of the two leading jets, $m_{jj}$, whereas in the boosted category the vectorial sum of the $\tau$ leptons and of the missing transverse momentum is considered as the second dimension.

\section{Background estimation methods and systematic uncertainties}

The Z$\rightarrow\tau\tau$ background is estimated from simulations. Corrections to the dilepton mass, dilepton $p_T$, and $m_{jj}$ for events with at least two jets, are derived from a pure Z$\rightarrow\mu\mu$ control region, and applied to the Z$\rightarrow\tau\tau$ background in the signal region. The normalization of the Z$\rightarrow\tau\tau$ background is also corrected on the basis of the data-to-prediction agreement in the Z$\rightarrow\mu\mu$ control region. Uncertainties in the yield of the Z$\rightarrow\tau\tau$ background go up to 7\%, and the uncertainties in the corrections are considered as shape uncertainties.\\

The W+jets background is estimated from simulations in the $e\mu$ and $\tau_h\tau_h$ final states. In the $e\tau_h$ and $\mu\tau_h$ channels, its yield is adjusted in such a way as to obtain perfect yield agreement in control regions defined by requiring the transverse mass between the light lepton and the transverse missing momentum to be above 80 GeV. These control regions are added to the fit to extract the results. The uncertainty in this procedure leads to uncertainties in the W+jets background yield up to 20\%.\\

The yield of the QCD multijet background in the $e\mu$, $e\tau_h$, and $\mu\tau_h$ final states is estimated by inverting the charge requirement on the $\tau$ leptons and subtracting other background contributions from the data. The yield is corrected to account for composition differences between the regions with same-sign or opposite-sign leptons. The QCD multijet background distributions are also extracted from a region with same-sign $\tau$ leptons, but the lepton isolation is in addition relaxed to improve the statistical precision. In the $\tau_h\tau_h$ final state, the QCD multijet background is derived from a region with opposite-sign $\tau_h$ candidates passing relaxed isolation conditions. Control regions composed of events with opposite-sign $\tau$ leptons passing inverted isolation criteria are added to the fit. The uncertainty in this procedure leads to uncertainties in the QCD multijet background yield up to 20\%.\\

Other backgrounds, including $t\bar{t}$, single top quark, or diboson production, are estimated from simulations. The yield of the $t\bar{t}$ background is corrected on the basis of the agreement between data and prediction in a control region obtained in the $e\mu$ final state by inverting the $p_{\zeta}$ requirement. Higgs boson decays to pairs to W bosons are considered as part of the background.\\

Uncertainties in the $\tau_h$ identification efficiency amount to 5\%, and 5\% is also considered for the $\tau_h$ trigger efficiency. The uncertainty in the $\tau_h$ energy scale is 1.2\% per decay mode before the fit, and it gets constrained to about one fourth of its initial value after the fit. Uncertainties in the $\tau_h$ energy scale and in the energy scale of the missing transverse momentum are the dominant uncertainties in the analysis. 

\section{Results}

\begin{figure}[htb]
\centering
\includegraphics[height=3in]{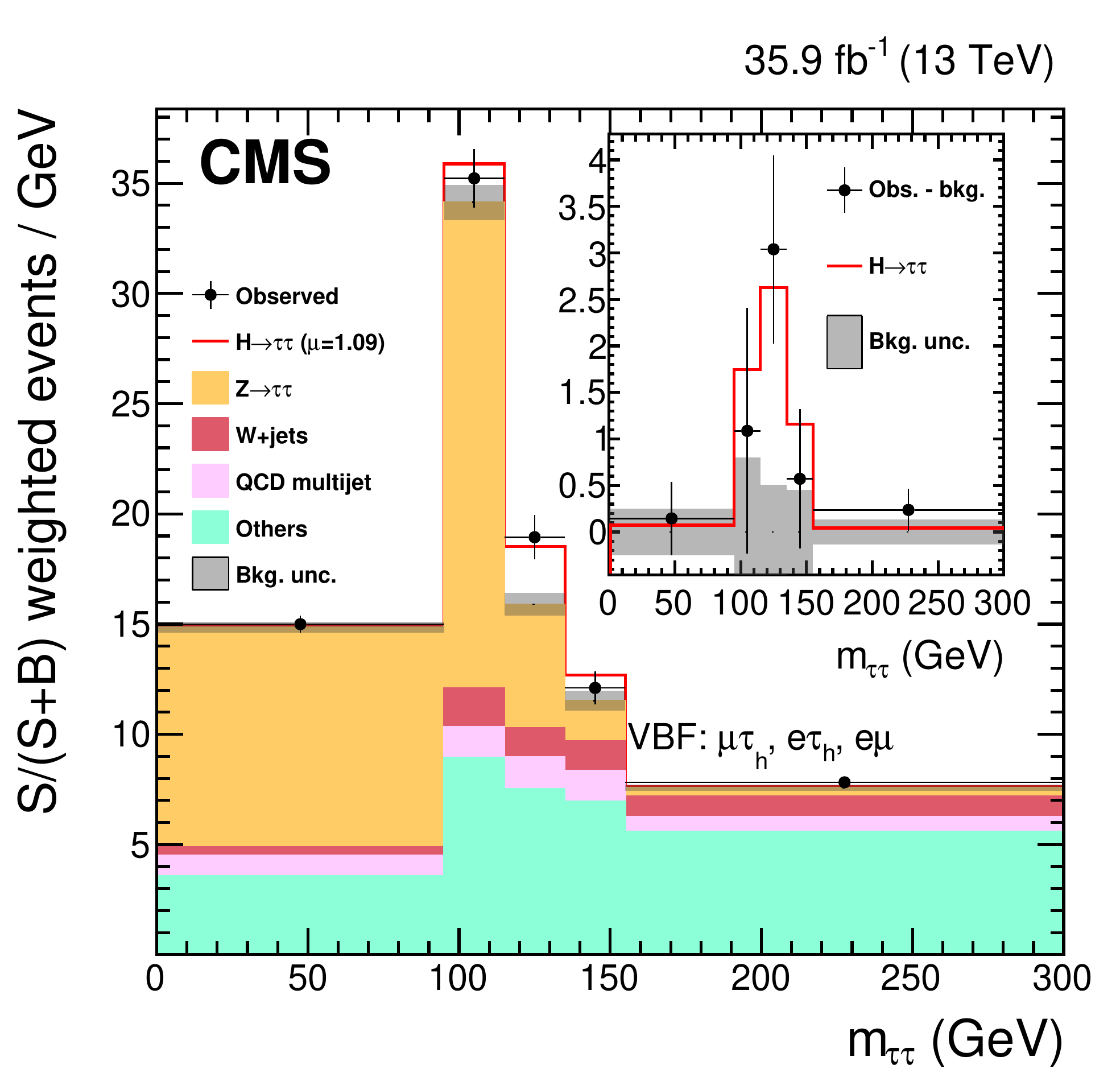}
\includegraphics[height=3in]{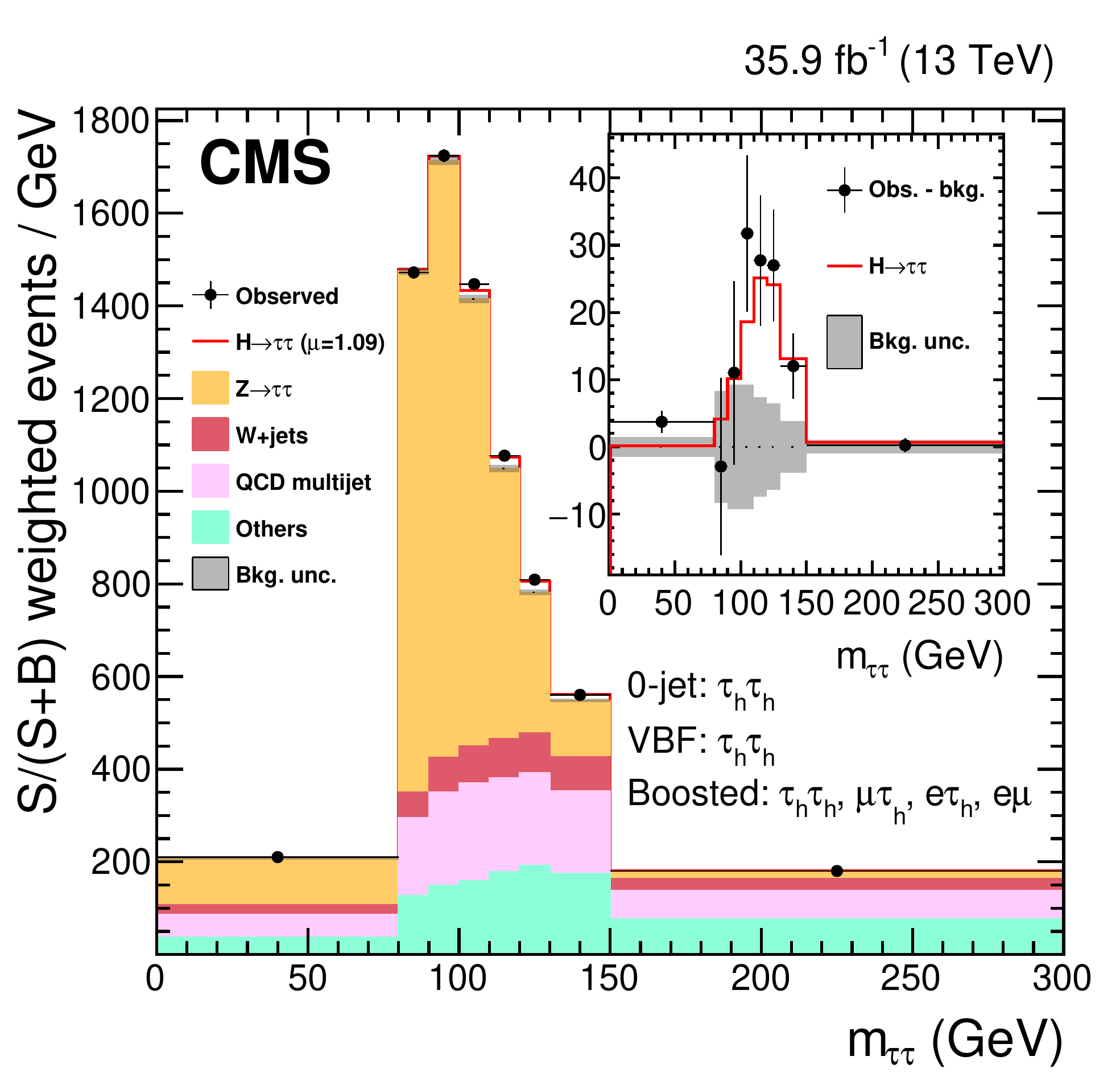}
\caption{ Combined observed and predicted $\tau\tau$ mass distributions. The left pane includes the VBF category of the $\mu\tau_h$, $e\tau_h$ and $e\mu$ channels, and the right pane includes other channels. The binning reflects the one used in the 2D distributions, and does not allow merging of the two figures. The normalization of the predicted background distributions corresponds to the result of the global fit, while the signal is normalized to its best fit signal strength. The mass distributions for a constant range of the second dimension of the signal distributions are weighted according to S/(S + B), where S and B are computed, respectively, as the signal or background contribution in the mass distribution excluding the first and last bins. The "Others" background contribution includes events from diboson, $t\bar{t}$, and single top quark production, as well as Higgs boson decay to a pair of W bosons and Z bosons decaying to a pair of light leptons. The background uncertainty band accounts for all sources of background uncertainty, systematic as well as statistical, after the global fit. The inset shows the corresponding difference between the observed data and expected background distributions, together with the signal expectation. The signal yield is not affected by the reweighting.}
\label{fig:figure1}
\end{figure}

The results are extracted with a binned likelihood fit to the 2D distributions in the signal region and to the distributions in the control regions. 
As shown in Fig.~\ref{fig:figure1}, an excess of events compatible with the SM Higgs boson is observed on top of the predicted backgrounds. The significance of the excess is 4.9 standard deviations, for 4.7 expected standard deviations.

\section{Summary}

 An observation of Higgs boson decays to $\tau$ leptons, based on data collected in pp collisions by the CMS detector in 2016 at a center-of-mass energy of 13 TeV, has been presented. The analysis targets both the gluon fusion and the vector boson fusion production mechanisms. The results are extracted via maximum likelihood fits in two-dimensional planes, and give an observed significance for Higgs boson decays to $\tau$ lepton pairs of 4.9 standard deviations, to be compared with an expected significance of 4.7 standard deviations. 
 This constitutes the most significant direct measurement of the coupling of the Higgs boson to $\tau$ leptons by a single experiment.


\begin{thebibliography}{99}


\bibitem{Aad:2012tfa} 
  G.~Aad {\it et al.}  [ATLAS Collaboration],
  Phys.\ Lett.\ B {\bf 716}, 1 (2012)
  [arXiv:1207.7214 [hep-ex]].
  
  
\bibitem{Chatrchyan:2012ufa} 
  S.~Chatrchyan {\it et al.}  [CMS Collaboration],
  Phys.\ Lett.\ B {\bf 716}, 30 (2012)
  [arXiv:1207.7235 [hep-ex]].
  
  \bibitem{HIG-16-043} 
  S.~Chatrchyan {\it et al.}  [CMS Collaboration],
  Submitted to Phys.\ Lett.\ B (2017)
  [arXiv:1708.00373 [hep-ex]].



\end{thebibliography}
\end{document}